\documentclass{article}
\usepackage{spconf,amsmath,graphicx}

\usepackage{cite}
\usepackage{enumitem}
\setlist{nosep, leftmargin=14pt}
\usepackage{amssymb}
\usepackage{mwe} % to get dummy images
\usepackage{multirow}
\usepackage{color}

\title{Inter-slice Super-resolution of Magnetic Resonance Images by Pre-training and Self-supervised Fine-tuning}
\name{\parbox{\linewidth}{\centering Xin Wang$^{1, \dagger}$ \qquad Zhiyun Song$^{1, \dagger}$ \qquad Yitao Zhu$^{2}$ \qquad Sheng Wang$^{1,4}$ \\ Lichi Zhang$^{1}$ \qquad Dinggang Shen$^{2,3,4}$ \qquad Qian Wang$^{2,3}$} % \textcolor{red}{Does ISBI require anonymous submission?}
\thanks{$\dagger$ Equal contribution to this work.}}
\address{
$^{1}$ School of Biomedical Engineering, Shanghai Jiao Tong University, Shanghai, 200030, China \\
$^{2}$ School of Biomedical Engineering \& State Key Laboratory of \\Advanced Medical Materials and Devices, ShanghaiTech University, Shanghai, 201210, China\\
$^{3}$ Shanghai Clinical Research and Trial Center, Shanghai, 201210, China\\
$^{4}$ Shanghai United Imaging Intelligence Co., Ltd., Shanghai, 200230, China\\
}
\begin{document}
%\ninept
%
\maketitle
In clinical practice, 2D magnetic resonance (MR) sequences are widely adopted. 
While individual 2D slices can be stacked to form a 3D volume, the relatively large slice spacing can pose challenges for both image visualization and subsequent analysis tasks, which often require isotropic voxel spacing.
To reduce slice spacing, deep-learning-based super-resolution techniques are widely investigated. 
However, most current solutions require a substantial number of paired high-resolution and low-resolution images for supervised training, which are typically unavailable in real-world scenarios.
In this work, we propose a self-supervised super-resolution framework for inter-slice super-resolution of MR images.
Our framework is first featured by pre-training on video dataset, as temporal correlation of videos is found beneficial for modeling the spatial relation among MR slices. 
Then, we use public high-quality MR dataset to fine-tune our pre-trained model, for enhancing awareness of our model to medical data.
Finally, given a target dataset at hand, we utilize self-supervised fine-tuning to further ensure our model works well with user-specific super-resolution tasks.
% that combines advantages of supervised pre-training on public dataset and self-supervised fine-tuning on clinical dataset. 
% Furthermore, we find that pre-training on large-scale video dataset enables higher performance for MRI super-resolution task. 
% Specifically, we pre-train our super-resolution model on video frame interpolation task, which adopts a similar idea with MRI slice interpolation of predicting middle frames (slices) based on givens frames (slices). 
The proposed method demonstrates superior performance compared to other self-supervised methods and also holds the potential to benefit various downstream applications.


\begin{figure*}[t]
\centering
\includegraphics[width=\linewidth]{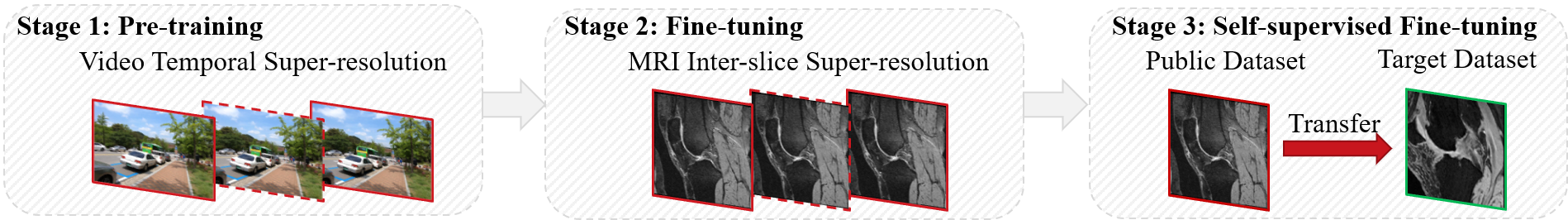}
\caption{The proposed three-stage super-resolution framework, combining advantages of pre-training, fine-tuning, and self-supervised fine-tuning.} \label{flowchart}
\end{figure*}

\section{Introduction}
% \textcolor{red}{Introduction is too long. Should be within one page.}
Magnetic resonance (MR) imaging is widely used for its non-invasive and detailed visualization of the human body. In clinical MR imaging, 2D sequences are widely employed for faster image acquisition, where multiple 2D slices can be stacked to create a 3D volume. 
Such volume typically has large inter-slice spacing, in contrast to the fine-grained intra-slice spacing~\cite{mri}.
The anisotropic voxel spacing in 3D can pose significant challenges for many automatic image processing software requiring isotropic voxel spacing as input.

One straightforward solution is to resample the volume. However, this operation can corrupt images by blurring or aliasing, particularly near the boundaries of organ/tissue with sharp intensity transitions.
Recently, several super-resolution (SR) techniques based on deep learning have been proposed \cite{ArSSR,DeepResolve,wang2022,SMORE,TSCNet,xuankai}. 
A common idea of these methods is to leverage the power of neural networks to learn a mapping from the low-resolution (\textit{LR}, with large inter-slice spacing) images to the high-resolution (\textit{HR}, with small inter-slice spacing) ones.

To establish the mapping, supervised learning is a straightforward paradigm \cite{ArSSR,DeepResolve,wang2022}.
% One can acquire real HR images, i.e., with near-isotropic voxels. 
% To create the HR-LR pairs for training, the HR images are downsampled to obtain the corresponding LR images. 
During training, the model takes an LR image as input and minimizes the discrepancy between the predicted and real HR image. 
However, the above paradigm faces the challenge that HR images sometimes cannot be acquired in clinical practice.
To avoid using real HR data for supervision, researchers have developed self-supervised solutions, which can be divided into two categories: resampling-based and synthesis-based methods. 
The resampling-based methods \cite{SMORE,TSCNet} first simulate images of lower resolution from original LR ones, learn a mapping from the lower-resolution images to the LR images, and then apply it to the LR images to predict the HR images. 
% Resample-based methods often results in inferior performance due to limited training data and the learning of indirect mapping. 
The synthesis-based methods \cite{xuankai} involve synthesizing HR images from the original LR images and training the SR model based on the synthesized images. 
% For example, \cite{xuankai} \emph{et al.}. propose to encode the neighboring slices into a low-dimensional space. Through interpolation within this latent space and decoding, more slices can be generated, which can then be stacked to create an HR volume with reduced slice spacing.
% However, the SR results of synthesis-based methods largely depends on the quality of synthesized images. The inherent uncertainty introduced by the generative process can significantly impact the performance of downstream SR task.
% \textcolor{red}{Xuankai's work is not in this category.}
% However, these existing methods all require a substantially large dataset to enable the self-supervised learning. Meanwhile, the learned models are difficult to generalize to different datasets and scenarios.

% the three methods listed above only use a few LR cases for training, so we can't say they require a substantially large dataset; 
% \textcolor{red}{There should be a paragraph to introduce recent progress of pre-training.}
The main drawbacks of current self-supervised SR methods lie in limited training data and indirect mapping. Specifically, the training set is derived from a few LR cases. And the SR mapping is learned from lower-resolution to low-resolution data, or from synthesized LR to synthesized HR data, rather than the real LR-HR pairs, which could significantly impede the model's performance.
One potential solution is to pre-train on extensive public datasets \cite{Med3D}, which offer strong initialization with high-quality data and alleviate the requirements for downstream data. 
% For example, Med3D \cite{Med3D}, following pre-training on eight diverse 3D segmentation datasets, demonstrates superior performance in the classification and segmentation of 3D medical images.
% Several studies \cite{Med3D, Liu_2023_ICCV} have employed an ensemble of public medical datasets to improve the model's generalization capabilities.
Considering that medical images are comparably difficult to collect, several studies \cite{rajpurkar2020appendixnet, videopretrain_ct} explored the use of video datasets as an alternative to medical images for capturing inter-slice correlation. 
% Nonetheless, it is still uncertain whether pre-training on external datasets will yield benefits for reconstruction tasks.

This paper presents a novel SR framework that combines the advantages of supervised pre-training on public data and self-supervised fine-tuning on user-specific data.
The pre-training offers high performance and generalization, and the fine-tuning further enables the model to adapt to specific data and task.
As depicted in Fig. \ref{flowchart}, the proposed framework consists of three stages: 
\textit{(1) Pre-training on video frame interpolation.}
It shares similarity with inter-slice SR of MR images as both tasks involve predicting intermediate frames/slices. Pre-training on abundant video data equips the model with strong prior for modeling the inter-slice correlation.
\textit{(2) Fine-tuning on high-quality MR dataset.}
The motivation is to adapt the pre-trained model to the domain of medical images. Thus, it becomes familiar with MR-specific context, enhancing its performance in MR image representation.
\textit{(3) Self-supervised fine-tuning on target dataset.}
Since the user-provided target dataset may have varying tissue types, modalities, or structural complexity, it is essential to adapt the model to user-specific cases through self-supervised fine-tuning.

The main contributions in this paper are listed as follows.
\begin{itemize}
\item We propose a framework for MR inter-slice SR that benefits from pre-training on abundant video data.
% combines the advantages of supervised training on publicly available datasets with the ability to be transferred to clinical scenarios.
\item We adapt the pre-trained SR model to user cases by self-supervised fine-tuning.
% The paper pioneers the study of pre-training on video frame interpolation task and demonstrates that sequential video then MRI pre-training outperforms just MRI pre-training .
\item We demonstrate superior SR performance over other self-supervised methods on knee MR images.
% The experimental results on multiple datasets demonstrate a significant improvement over other self-supervised SR methods. 
\end{itemize}

\section{Methodology}
Our method contains three parts: (1) pre-training on video frame interpolation, (2) fine-tuning on high-quality MR dataset for transferring the pre-trained model to MR images, and (3) self-supervised fine-tuning on target dataset for further transferring to user cases. 
% \textcolor{red}{Motivations or aims of three stages need be introduced, either here or in the final of Introduction.}
% We will introduce individual parts in subsequent sections.

\subsection{Pre-training on Video Frame Interpolation}
%  large-scale video frame interpolation dataset continuous slice representation
Video pre-training is found to be beneficial for 3D medical tasks \cite{ke2023video}.
Although numerous pre-trained video models are publicly available, most are specialized for discriminative tasks, such as video classification and action recognition. 
Consequently, they may not effectively capture fine-grained features required by generative tasks. 
In our case, we find that directly deploying a public pre-trained video classification model yields no improvement in the performance of downstream SR task.
Therefore, we first pre-train the SR model on video frame interpolation task. 

We adopt SA-INR \cite{wang2022} as the SR model, which parameterizes a frame or slice sequence as a continuous function of spatial coordinates. % so that the middle frames or slices can be reconstructed by querying all the required coordinates. 
The training process is illustrated as follows.
First, we randomly choose a sub-sequence $I_{seq}$ with a length of $15n+1$, retaining one frame for every $n$ and taking the rest as ground truth.
Here $n$ denotes the down-sampling factor.
Given two adjacent frames $I^i,I^{i+n}\in I_{seq}$, we synthesize an intermediate frame $I^{i+k}$ ($0 \le k \le n$). To achieve this goal, we input $I^i$ and $I^{i+n}$, as well as the spatial coordinates $C^{i+k}$ of frame $I^{i+k}$, to SA-INR, which will return a predicted intermediate frame $\hat{I}^{i+k}$.
Finally, we calculate the $L_1$ loss to enforce the pixel-wise consistency between the predicted frame $\hat{I}^{i+k}$ and the real intermediate frame $I^{i+k}$:
\begin{align*}
    L&=\Vert{\hat{I}^{i+k}-I^{i+k}}\Vert \\
     &=\Vert{\mathcal{F}(I^i,I^{i+n},C^{i+k})-I^{i+k}}\Vert,
\end{align*}   
where $\mathcal{F}(\cdot)$ denotes the mapping function of the SR network.

By reconstructing the missing temporal frames, the model not only effectively captures fine-grained visual cues, but also learns spatio-temporal information in the video that mimics inter-slice relation in an MR volume.
% In our implementation, we pre-train the model on the large-scale and high-quality REDS-VTSR dataset \cite{REDS_VTSR}, which consists of 270 long video sequences. Our experiments reveal that long sequences are more conducive to our task since they contain richer temporal information.
% the above paragraph is moved to dataset introduction paragraph
% \textcolor{red}{In 2.2, you have math formulation for the fine-tuning task. The same math terms should be used for formulation of pre-trainng here. In this way, reviewers can understand the connection among all three stages.}

\subsection{Fine-tuning on High-quality MR Dataset}
In the second stage, we transfer the SR model pre-trained on the video dataset to the MR domain. This fine-tuning enables the model to capture the essential anatomical characteristics by learning from high-quality MR images. 
Note that the high-quality MR images here are relatively costly to acquire, thus NOT widely used in clinical practice. We utilize these images to supervise the fine-tuning of the SR model.
% This step acts as a crucial bridge between the pre-training phase and the ultimate application in user-specific cases.
% The goal of the first two stages is to equip the model with powerful abilities of representation learning and slice prediction.
% Consider an LR volume $I_{LR}$ with spatial resolution of $a\times a \times c$ where $c \ge a$, the aim of the SR model is to restore the HR volume $I_{HR}$ with spatial resolution of $a\times a \times a$.
% \textcolor{red}{For the next, I didn't see any imformation about fine-tuning. You didn't mention your pre-trained model. How to c onnect your fine-tuning to pre-trained model?}

After pre-training the SR model on the video frame interpolation task, we fine-tune all the parameters of the model on the MR inter-slice SR task.
Specifically, for each iteration of training, we randomly select an HR volume $I_{HR}$ with spatial resolution of $a\times a \times a$ from the training set. Then we downsample the volume along the $z$-axis and obtain an LR volume $I_{LR}$ with spatial resolution of $a\times a \times na$  where $n$ represents the downsampling factor. Next, we extract two adjacent slices $S^i_{x,y}$ and $S^{i+n}_{x,y}$ from $I_{HR}$ as the input of SA-INR, and reconstruct the desired middle slice $\hat{S}^{i+k}_{x,y}$ ($0 \le k \le n$). 
Finally, we calculate and minimize the loss between $\hat{S}^{i+k}_{x,y}$ and the real intermediate slice $S^{i+k}_{x,y}$.
% \begin{align*}
%     L&=\Vert{\hat{S}^{i+k}_{x,y}-S^{i+k}_{x,y}}\Vert \\
%      &=\Vert{\mathcal{F}(S^i_{x,y},S^{i+1}_{x,y},Z^{i+k}_{x,y})-S^{i+k}_{x,y}}\Vert
% \end{align*}   
% where $\mathcal{F}(\cdot)$ denotes the mapping function of the SR network.

\subsection{Self-supervised Fine-tuning on Target Dataset}
% \textcolor{red}{A basic question is why we need this step. A second question is how it connects with previous two steps.}
Finally, we conduct self-supervised fine-tuning on the target dataset. This step is essential because directly applying the SR model trained on public dataset to specific user cases may lead to performance decrease due to domain gap.
Regarding fine-tuning, the difference between Stage 2 and Stage 3 is that only LR images are available to fine-tune Stage 3, while real HR images are available to supervise Stage 2.
Thus, to acquire HR-LR pairs for Stage 3, we further downsample the LR images in the target dataset.
Specifically, to enable tailored solution for each single subject, we conduct subject-based fine-tuning here. 
Given an LR volume $I_{LR}$ with spatial resolution of $a\times a \times c$ ($c \ge a$), we downsample the volume along $x$-axis (or $y$-axis) to acquire  $I_{LR,x \downarrow}$ with spatial resolution of $n a\times a \times c$, where $n$ represents the downsampling factor. In this way, we can build a training set $\{I_{LR}, I_{LR,x \downarrow}\}$.
% During training, we input two adjacent slices $S^i_{y,z}$ and $S^{i+1}_{y,z}$ from $I_{LR,x \downarrow}$ and let model predict the intermediate slice $\hat{S}^{i+k}_{y,z}$ ($0 \le k \le 1$). Then we calculate the loss between $\hat{S}^{i+k}_{y,z}$ and the real intermediate slice $S^{i+k}_{y,z}$.

We investigate several fine-tuning strategies, including fine-tuning all parameters, freezing a subset of parameters, and employing the parameter-efficient fine-tuning technique (PEFT) \cite{hu2021lora}. Our experiments reveal that fine-tuning all parameters (2.1M) is the most effective and does not impose significant computational resource burden.
% mem: 3.507GB, GPUmem: 22.719GB
% model: #params=2.1M
The subject-based fine-tuning takes about 1 minute on an A100 40G card, and the subsequent SR process takes about 20 seconds.
% \textcolor{red}{This sentence is very confusing to me.}
% By fine-tuning the model using the target subject, the model adapts better to the specific characteristics and nuances of the subject, leading to improved accuracy and effectiveness in real-world applications.

\begin{table}[t]
\centering
\caption{\label{tab:quantitative_result}
Quantitative results for different self-supervised super-resolution methods on the SKI10 dataset.}
\begin{tabular}{lccccc}
\hline
\multirow{2}{*}{Method} & \multicolumn{2}{c}{PSNR} &  & \multicolumn{2}{c}{SSIM} \\ \cline{2-3} \cline{5-6} 
 & Mean & SD &  & Mean & SD \\ \hline
Trilinear & 28.26 & 2.59 &  & 0.8108 & 0.0468 \\
TSCNet \cite{TSCNet} & 27.37 & 2.46 &  & 0.7841 & 0.0478 \\
SMORE \cite{SMORE} & 29.33 & 2.76 &  & 0.8450 & 0.0434 \\
Xuan \emph{et al.} \cite{xuankai} & 30.39 & 2.82 &  & 0.8425 & 0.0471 \\
Proposed & \textbf{30.88} & 2.83 &  & \textbf{0.8517} & 0.0512 \\ \hline
Supervised & 31.22 & 2.83 &  & 0.8628 & 0.0471 \\
\hline
\end{tabular}
\end{table}

\begin{table}[t]
\centering
\caption{\label{tab:ablation_result}
Quantitative results for the ablation study. VP denotes pre-training on video, SF denotes supervised fine-tuning on MRI dataset, and SSF denotes self-supervised fine-tuning on target dataset.}
\begin{tabular}{cccccccc}
\hline
\multirow{2}{*}{VP} & \multirow{2}{*}{SF} & \multirow{2}{*}{SSF} & \multicolumn{2}{c}{PSNR} &  & \multicolumn{2}{c}{SSIM} \\ \cline{4-5} \cline{7-8} 
 &  &  & Mean & SD &  & Mean & SD \\ \hline
 & \checkmark &  & 30.19 & 2.77 &  & 0.8216 & 0.0591 \\
  & \checkmark & \checkmark & 30.40 & 2.78 &  & 0.8342 & 0.0569 \\
\checkmark & \checkmark & & 30.73 & 2.80 &  & 0.8404 & 0.0540 \\
\checkmark & \checkmark & \checkmark & \textbf{30.88} & 2.83 &  & \textbf{0.8517} & 0.0512 \\
\hline
\end{tabular}
\end{table}

\section{Experiments}

\subsection{Datasets and Experimental Setup}
\textbf{Video Dataset}:
We use the REDS-VTSR dataset \cite{REDS_VTSR} for pre-training, which consists of 270 sequences with a frame rate of 60fps. Each sequence contains 180 frames with a size of 720px$\times$1280px. 
We have found that long sequences are more conducive to our task since they contain richer temporal information.
During training, we transform each RGB frame to gray, resize it to 90px$\times$160px and then crop a 64px$\times$64px patch.
We pre-train our model on the REDS-VTSR dataset for 1000 epochs with an initial learning rate of $1e^{-4}$, which is halved every 200 epochs.

\noindent \textbf{Public MR Dataset}:
% We adapt the pre-trained video SR model to the scenario of knee MR imaging, by using the publicly available Osteoarthritis Initiative (OAI) dataset \cite{OAI}.
We adapt the pre-trained video SR model to the publicly available Osteoarthritis Initiative (OAI) dataset \cite{OAI}.
Specifically, we collect a total of 350 cases (3D DESS, spatial resolution: 0.3646mm $\times$ 0.3646mm $\times$ 0.7mm), where 300 cases are used for training and 50 cases for validation.
During training, we randomly crop an HR patch with a size of $64\times64\times16n$ for each case, from which we simulate an LR patch ($64\times64\times16$) following \cite{iglesias2021joint}. 
We fine-tune the model on the OAI dataset for 1000 epochs, using the same experimental settings as in the pre-training phase.

\noindent \textbf{Target MR Dataset}:
% While the previous public MR dataset has HR ground truth, more clinical datasets do not include HR cases. 
We perform self-supervised fine-tuning on the Segmentation of Knee Images 2010 (SKI10) dataset \cite{SKI10}.
We collect 150 cases (100 for training, 50 for testing) with T1 or T2-weighted modalities
% , where only the test set is used in our experiment. 
% Note that the sequences and image appearances of SKI10 are different from OAI.
All images are scanned in the sagittal plane with the spatial resolution of 0.4mm $\times $ 0.4mm $\times $ 1.0mm.
We downsample them by 4 times to simulate LR data with resolution of 0.4mm $\times $ 0.4mm $\times $ 4.0mm.
% while the HR ones can be used for quantitative evaluation.
% During fine-tuning, we further downsample the LR cases along the axial axis to simulate LR-LR$\downarrow$ pairs. 
Note that the supervised method is trained using the training set of SKI10 and evaluated on the testing set. In contrast, the self-supervised methods are trained and evaluated solely on the testing set, without referencing to any HR images.
For each case, we randomly extract 100 patches and fine-tune the SR model for 5 epochs.

\noindent \textbf{Comparison Methods}:
We compare our method with three self-supervised SR methods: (1) TSCNet \cite{TSCNet}, a two-stage method that first initiates the network by through-plane LR-HR pairs, and then refines the network using cyclic-based interpolation;
% As the method only performs $\times$2 super-resolution task, we iteratively perform the task to achieve larger scales (e.g., $\times$4).  
(2) SMORE \cite{SMORE}, which first degrades sagittal slices and trains a 2D neural network on sagittal pairs, and then applies it to reconstruct HR coronal and axial views;
(3) Xuan \emph{et al.} \cite{xuankai}, which synthesizes HR image using the variational auto-encoder \cite{vae}, and trains a super-resolution network based on these synthesized pairs.

\subsection{Comparative Results on SKI10}
% We simulate the LR images by degrading the HR image in the SKI10 dataset with a downsampling factor of 4, rendering MR images with the spatial resolution of 0.4mm $\times $ 0.4mm $\times $ 4.0mm.
% All the self-supervised methods are trained on the simulated LR data without referring to the HR images.

\begin{figure}[ht]
\includegraphics[width=\columnwidth]{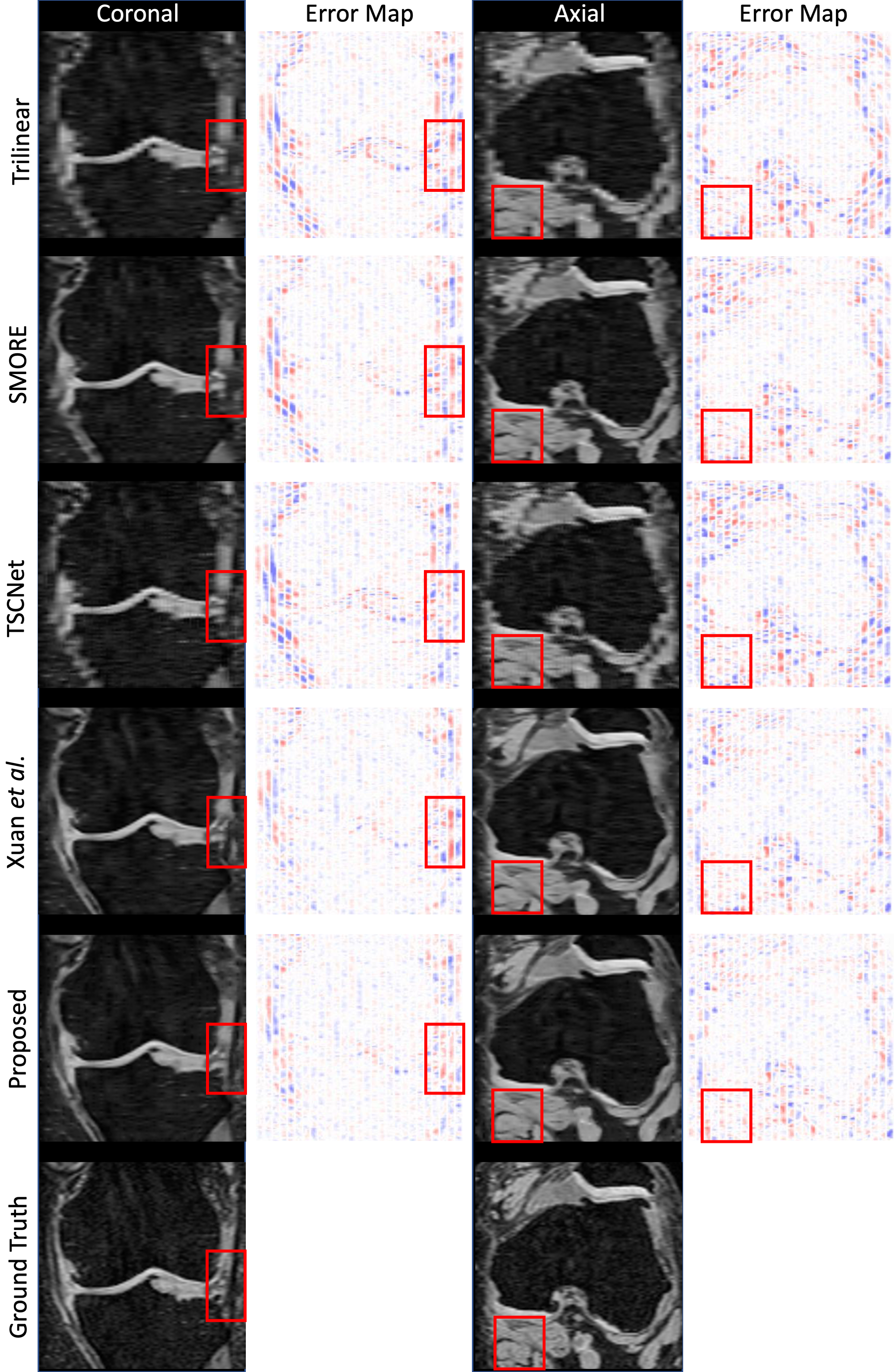}
\caption{Qualitative results and error maps for different self-supervised SR methods on the SKI10 dataset.} \label{qualitative_result}
\end{figure}

We conduct quantitative evaluation using two commonly used metrics, \emph{i.e.,} the Peak Signal-to-Noise Ratio (PSNR) and Structural Similarity (SSIM) \cite{metrics}, and also provide the qualitative results.
As reported in Table \ref{tab:quantitative_result}, almost all of the self-supervised methods outperform the baseline of trilinear resampling, except for TSCNet.
% which targets on 2$\times$ super-resolution and might not be suitable for the task. 
We also observe the impressive PSNR produced by Xuan \emph{et al.} However, the SSIM results produced by Xuan \emph{et al.} are inferior, as there are some distorted structures, as confirmed by red boxes in Fig. \ref{qualitative_result}.
Instead, our method reconstructs the most reliable MR images.
By leveraging the combined benefits of supervised pre-training and self-supervised fine-tuning, the resulting images exhibit superior quality, showcasing enhanced inter-slice continuity and fidelity.
We also perform the supervised SR on the training set of SKI10 and report the metrics on the test set.
It is inspiring to find that our method is close to this upper bound, with a margin of only 0.34dB in terms of PSNR. 

\subsection{Ablation Study}
We evaluate the impact of Video Pre-training (VP) and Self-supervised Fine-tuning (SSF) by combining each of them with Supervised Fine-tuning (SF) on the same dataset.
As reported in Table \ref{tab:ablation_result}, some important findings are observed. % we observe that each of the strategy is beneficial for the task.
% The results are reported in Table \ref{tab:ablation_result}.

\subsubsection{Video Pre-training Benefits MRI Super-Resolution}
To evaluate the benefits brought by video pre-training, we start with supervised training on OAI. 
As denoted by the VP+SF row in Table \ref{tab:ablation_result}, there is a significant performance improvement when using video pre-training to initialize the SR model.
This indicates that video pre-training can benefit downstream MR inter-slice SR task, probably due to the following two reasons. 
First, video frames are sequential data in nature and exhibit temporal consistency; pre-training on video frame interpolation task helps the SR model capture and leverage this temporal information, which is particularly beneficial when dealing with MR volume, where inter-slice connection is crucial. Second, by pre-training from abundant video data, an SR model can learn to extract a wealth of visual information, which not only enhances the quality of MR images, but also accelerates the convergence of the SR model during fine-tuning on MR data, reducing the need for extensive training on the medical data.

\subsubsection{Self-supervised Fine-tuning Ensures the Adaptation to Specific Cases}
Given the intricate structural and modal nuances of MR images, the models trained on specific public datasets may face challenges when generalizing to complex cases during evaluation. As shown in Table \ref{tab:ablation_result}, the performance of supervised training on the OAI dataset is even inferior to most of the self-supervised training methods on the target SKI10 dataset, particularly in terms of SSIM.
After performing the subject-based self-supervised fine-tuning, the SR model successfully adapts to the target dataset and yields significantly superior results.
Moreover, the best results are achieved when combining the two strategies (VP and SSF) together, as denoted by the VP+SF+SSF row.

\section{Conclusion}

In summary, our three-stage self-supervised  framework offers a compelling solution to the challenge of implementing SR in clinical settings where HR data is absent. By combining supervised pre-training on high-quality dataset and self-supervised fine-tuning on target dataset, we achieve superior results compared to state-of-the-art methods.

Meanwhile, we demonstrate the effectiveness of video pre-training for MR modeling, bridging the task gap between 
video temporal SR and MR inter-slice SR. Given the scarcity of medical data in contrast to the abundance of video data, there is a promising potential for improving 3D medical task through video pre-training.

\section{Compliance with Ethical Standards}
This research study was conducted retrospectively using human subject data made available in open access. Ethical approval was not required as confirmed by the license attached with the open access data.

\bibliographystyle{IEEEbib}
\bibliography{strings,refs}

\end{document}